\documentclass[12pt]{article}
\usepackage{latexsym}
\usepackage{epsfig}
{\catcode `\@=11 \global\let\AddToReset=\@addtoreset}
\AddToReset{equation}{section}
\renewcommand{\theequation}{\thesection.\arabic{equation}}  

\def\greaterthansquiggle{\raise.3ex\hbox{$>$\kern-.75em\lower1ex\hbox{$\sim$}}}
\def\lessthansquiggle{\raise.3ex\hbox{$<$\kern-.75em\lower1ex\hbox{$\sim$}}}
\newcommand{\beq}{\begin{equation}}
\newcommand{\eeq}{\end{equation}}
\newcommand{\beqa}{\begin{eqnarray}}
\newcommand{\eeqa}{\end{eqnarray}}
\newcommand{\beqan}{\begin{eqnarray*}}
\newcommand{\eeqan}{\end{eqnarray*}}
\newcommand{\ba}{\begin{array}}
\newcommand{\ea}{\end{array}}
\newcommand{\no}{\nonumber}

\newcommand{\ra}{\rightarrow}

\newcommand{\ve}{\varepsilon}
\newcommand{\vp}{\varphi}

\newcommand{\dg}{\dagger}
\newcommand{\wt}{\widetilde}
\newcommand{\wh}{\widehat}

\newcommand{\cL}{{\cal L}}
\newcommand{\M}{{\cal M}}

\newcommand{\cO}{{\cal O}}

\newcommand{\Q}{{\cal Q}}

\newcommand{\st}{\stackrel}
\newcommand{\dfrac}{\displaystyle \frac}

\newcommand{\nn}{\nonumber \\}
\newcommand{\bea}{\begin{eqnarray}}
\newcommand{\eea}{\end{eqnarray}}

\normalsize
\begin{document}
\bibliographystyle{plain}
\begin{titlepage}
\begin{flushright}
UWThPh-2000-17\\
LNF-00/016(P)\\
FTUV/00-0614\\
IFIC/00-29\\
June 2000
\end{flushright}
\vspace{2.5cm}
\begin{center}
{\Large \bf Electromagnetism in\\ Nonleptonic Weak Interactions*}
\\[40pt]
G. Ecker$^{1}$, G. Isidori$^{2}$, G. M\"uller$^{1,\#}$, H. Neufeld$^{1,3}$
and 
A. Pich$^{3}$

\vspace{1cm} 
${}^{1)}$ Institut f\"ur Theoretische Physik, Universit\"at 
Wien\\ Boltzmanngasse 5, A-1090 Wien, Austria \\[10pt] 
 
${}^{2)}$ INFN, Laboratori Nazionali di Frascati, P.O. Box 13, I-00044
Frascati, Italy \\[10pt]

${}^{3)}$ Departament de F\'{\i}sica Te\`orica, IFIC, Universitat de 
Val\`encia - CSIC\\ 
Apt. Correus 2085, E-46071 Val\`encia, Spain

\vfill
{\bf Abstract} \\
\end{center}
\noindent
We construct a low-energy effective field theory that permits the
complete treatment of isospin-breaking effects in nonleptonic weak
interactions to next-to-leading order. To this end, we enlarge the 
chiral Lagrangian describing strong and $\Delta S =1$ weak
interactions by including electromagnetic terms with the photon 
as additional dynamical degree of freedom.
The complete and minimal list of local terms at next-to-leading order 
is given. We perform the one-loop renormalization at the level of the 
generating functional and specialize to $K \to \pi\pi$ decays.

\vfill
 
\vfill
\noindent * Work supported in part by  TMR, EC-Contract No. ERBFMRX-CT980169 
(EURODA$\Phi$NE) and by DGESIC (Spain) under grant No. PB97-1261. \\
\noindent \# Present address: Jagdweg 23, D-53115 Bonn, Germany

\end{titlepage}
\section{Introduction}
\label{sec:Introduction}
\renewcommand{\theequation}{\arabic{section}.\arabic{equation}}
\setcounter{equation}{0}

Although isospin violation in nonleptonic weak interactions has been
investigated many times in the past systematic treatments have
appeared only rather recently \cite{EdR89,CDG1,CDG2,CDG3,WM99}. 
The topic is both of general interest and of considerable
phenomenological relevance. Precise determinations of weak decay
amplitudes are needed for many purposes, in particular for a
reliable calculation of CP violation in the $K^0-\bar{K^0}$ system. 
In the standard model, isospin violation arises from the quark mass 
difference $m_u-m_d$ and from electromagnetic corrections. Although
these effects are expected to be small in general they are 
amplified in nonleptonic weak transitions. Because of the suppression 
of amplitudes with $\Delta I > 1/2$, isospin violation in the 
dominant $\Delta I=1/2$ amplitudes leads to significantly enhanced
corrections for the sub-dominant amplitudes. In fact, a quantitative
analysis of the $\Delta I=1/2$ rule is only possible with the
inclusion of isospin-violating effects.

At first order in a systematic low-energy expansion, isospin breaking
in the leading octet amplitudes of nonleptonic kaon decays is of
order $G_8 (m_u-m_d)$ and $G_8 e^2$ where $G_8$ denotes the strength
of the effective octet coupling. The corrections appear in the mass 
differences of charged and neutral mesons, via $\pi^0-\eta$ mixing  
and through electromagnetic penguins \cite{BW84} in the effective 
nonleptonic weak Hamiltonian. However, there are
good reasons to believe that the problem cannot be understood at
lowest order only. For instance, the resulting (tree-level)
corrections do not produce a $\Delta I=5/2$ component for which there
is some phenomenological evidence \cite{EdR89}.

The chiral realization of isospin violation due to the light quark
mass difference is available also at next-to-leading order. The
purpose of this paper is to close the gap in the electromagnetic
sector by
\begin{itemize} 
\item completing the construction of the effective chiral Lagrangian
of $\cO(G_8 e^2 p^2)$ and
\item performing the complete renormalization at the one-loop level
for nonleptonic weak transitions including electromagnetic corrections.
\end{itemize} 
As our notation indicates, we only consider corrections to the leading
octet part of the nonleptonic weak Hamiltonian. The results are 
applicable to the analysis of both $K\to 2 \pi$ 
\cite{CDG1,CDG2,CDG3,EIMNP2} and $K \to 3 \pi$ decays.

We start in Sec.~\ref{sec:symm} by recalling the ingredients for the
construction of effective theories of strong, electromagnetic and
nonleptonic weak interactions. In Sec.~\ref{sec:Leff} we review the
effective Lagrangian of lowest order. For this Lagrangian, we
evaluate the one-loop divergence functional by standard
heat-kernel techniques in Sec.~\ref{sec:divs}. The new parts are terms 
of $\cO(G_8 e^2 p^2)$
which arise also from using the equations of motion to transform to
the standard bases for the nonleptonic weak Lagrangian of $\cO(G_8 p^4)$
\cite{EKW93} and for the electromagnetic Lagrangian of $\cO(e^2 p^2)$
\cite{Urech}. In the following section we construct the complete and
minimal Lagrangian of $\cO(G_8 e^2 p^2)$ making use of CPS 
symmetry \cite{CPS}, Cayley-Hamilton relations, partial
integration in the action and of the equations of motion. We order
the terms in the effective Lagrangian according to their physical
relevance: $K \to \pi\pi$ amplitudes receive contributions from the 
first 12 operators, the next two appear in $K \to 3 \pi$ and the
rest turns out not to be relevant phenomenologically.
In Sec.~\ref{sec:decays}, we present the divergences for
the three $K \to 2 \pi$ amplitudes and compare with the results of
direct one-loop calculations \cite{CDG2,EIMNP2}. We summarize
our findings in Sec.~\ref{sec:Conclusions}. Various quantities
appearing in the heat-kernel expansion of the generating functional
are collected in the Appendix.

\section{Symmetries}
\label{sec:symm} 
\renewcommand{\theequation}{\arabic{section}.\arabic{equation}}
\setcounter{equation}{0}
For a complete treatment of isospin-breaking effects in nonleptonic
kaon decays, an appropriate effective Lagrangian with the pseudoscalar octet 
and the photon as dynamical degrees of freedom has to be used. The symmetries 
of the standard model are serving as the basic guiding principles for its
construction. Starting with QCD 
in the chiral limit $m_u = m_d = m_s = 0$, the resulting symmetry
under the chiral group $G = SU(3)_L \times SU(3)_R$ is spontaneously
broken to $SU(3)_V$. The pseudoscalar mesons $(\pi,K,\eta)$ are interpreted
as the corresponding Goldstone fields $\vp_i$ ($i = 1,\ldots,8$) acting
as coordinates of the coset space $SU(3)_L \times SU(3)_R/SU(3)_V$.
The coset variables $u_{L,R}(\vp)$ are transforming as
\beqa
u_L(\vp) &\st{G}{\ra}& g_L u_L(\vp) h(g,\vp)^{-1}~, \no \\
u_R(\vp) &\st{G}{\ra}& g_R u_R(\vp) h(g,\vp)^{-1}~, \no \\
g = (g_L,g_R) &\in& SU(3)_L \times SU(3)_R~,
\eeqa
where $h(g,\vp)$ is the nonlinear realization of $G$ \cite{CCWZ}.

The photon field $A_\mu$ is introduced in
\beq \label{photcoupl}
u_\mu = i [u_R^\dg (\partial_\mu - i r_\mu)u_R - u_L^\dg
(\partial_\mu - i l_\mu)u_L]
\eeq
by adding appropriate terms to the usual external vector and axial-vector
sources $v_\mu$, $a_\mu$:
\beqa \label{sources}
l_\mu &=& v_\mu - a_\mu - e Q_L A_\mu ~, \no \\
r_\mu &=& v_\mu + a_\mu - e Q_R A_\mu~.
\eeqa
The $3 \times 3$ matrices $Q_{L,R}$ are 
spurion fields with the transformation properties
\beq
Q_L \st{G}{\ra} g_L Q_L g_L^\dg~, \qquad
Q_R \st{G}{\ra} g_R Q_R g_R^\dg
\eeq
under the chiral group. We also define
\beq \label{Qhom}
\Q_L := u_L^\dg Q_L u_L~, \qquad
\Q_R := u_R^\dg Q_R u_R
\eeq
transforming as
\beqa
\Q_L &\st{G}{\ra}& h(g,\vp) \Q_L h(g,\vp)^{-1}~, 
\no \\
\Q_R &\st{G}{\ra}& h(g,\vp) \Q_R h(g,\vp)^{-1}~.
\label{homtr}
\eeqa
At the end, one identifies $Q_{L,R}$ with the quark charge matrix
\beq \label{Qem}
Q = \left[ \ba{ccc} 2/3 & 0 & 0 \\ 0 & -1/3 & 0 \\ 0 & 0 & -1/3 \ea
\right].
\eeq

External scalar and pseudoscalar sources are combined in 
\beq \label{chi}
\chi = s + i p ~.
\eeq
For the construction of the effective Lagrangian, it is convenient to introduce 
the quantities
\beq \label{chipm}
\chi_{\pm} = u_R^\dg \chi u_L \pm u_L^\dg \chi^\dg u_R
\eeq
with the same chiral transformation properties as $\Q_L, \Q_R$
in (\ref{homtr}).

After integrating out the heavy degrees of freedom, the $\Delta S = 1$ weak 
interactions can be described in terms of an effective four-fermion 
Hamiltonian \cite{OPE}.
With respect to  the chiral group $G$, this effective Hamiltonian
transforms as the direct sum
\beq
(8_L, 1_R) + (27_L, 1_R) + (8_L,8_R)~,
\eeq
where the first piece, contributing only to $\Delta I = \frac{1}{2}$ 
transitions, is largely dominant. In this work we shall consider 
only the electromagnetic corrections induced by 
the dominant octet part of the effective Hamiltonian. To this end we
introduce a weak spurion $\lambda$ that is finally taken at 
\begin{equation} \label{lambda}
\lambda = \frac{\lambda_6 - i \lambda_7} {2}
= \left[ \ba{ccc} 0 & 0 & 0 \\ 0 & 0 & 0 \\ 0 & 1 & 0 \ea
\right] ,
\eeq
where $\lambda_{6,7}$ are Gell-Mann matrices. In analogy to
(\ref{Qhom}) we also define
\beq \label{Delta}
\Delta:= u_L^\dg \lambda u_L~,
\eeq
transforming again as in (\ref{homtr}) under chiral transformations.

Although CP is broken by the weak interactions, the 
$\Delta S = 1$ transitions
are still invariant under the so-called CPS symmetry \cite{CPS}: a CP 
transformation followed by a subsequent interchange of $d$ and $s$
quarks. This symmetry is also obeyed by strong and electromagnetic 
interactions, provided the 2-3 indices of the
external fields are also exchanged appropriately
(this implies, in particular, the exchange $m_s \leftrightarrow  m_d$
in the mass terms).
The explicit CPS transformation properties of the several building blocks 
introduced so far are given by
\beqa \label{CPStransf}
u_{\mu}(x) &\st{CPS}{\ra}& - \epsilon(\mu) S u_\mu^T(\wt{x}) S~, \no \\
\chi_{\pm} (x) &\st{CPS}{\ra}& \pm S \chi_{\pm}^T(\wt{x}) S~, \no \\
\Q_{L,R} (x) &\st{CPS}{\ra}& S \Q_{L,R}^T(\wt{x}) S~, \no \\
\Delta (x) &\st{CPS}{\ra}& S \Delta^T(\wt{x}) S~, 
\eeqa
with
\beq
\wt{x} = (x^0,-\st{\ra}{x})~, \quad \epsilon(0) = 1~, 
\, \epsilon(1)=\epsilon(2)=\epsilon(3)=-1~,
\eeq
and
\beq \label{S}
S = \left[ \ba{ccc} 1 & 0 & 0 \\ 0 & 0 & 1 \\ 0 & 1 & 0 \ea
\right].
\eeq

\section{The effective Lagrangian at lowest order}
\label{sec:Leff} 
\renewcommand{\theequation}{\arabic{section}.\arabic{equation}}
\setcounter{equation}{0}

With the building blocks introduced in the previous section we may now assemble
our effective Lagrangian.
We adopt an expansion scheme where the n-th order is related to terms
of order $p^n$ in the strong and weak sector and to terms of order $e^2 p^{n-2}$
in the electromagnetic sector where $p$ denotes a typical meson momentum. 
Terms of $\cO (e^4)$ will be neglected throughout.

To lowest order $(n=2)$, 
our effective theory consists of the following parts: the strong sector is  
represented by the nonlinear sigma model in the presence of the external
sources $v_\mu, a_\mu, \chi$ \cite{GL85} and the photon coupling
introduced in (\ref{sources}):
\begin{equation} \label{L2strong}
\frac{F^2}{4} \; \langle u_\mu u^\mu + \chi_+\rangle ~.
\eeq
The symbol $\langle \;\rangle$ denotes the trace in three-dimensional flavour
space and $F$ is the pion decay constant in the chiral limit. 
Explicit chiral symmetry breaking by the non-vanishing masses of the 
light quarks is achieved by evaluating the generating functional at 
\beq \label{Mquark}
\chi = 2 B \M_{\rm quark}
 =  2 B \left[ \ba{ccc} m_u & 0 & 0 \\ 0 & m_d & 0 \\ 0 & 0 & m_s \ea
\right].
\eeq
The quantity $B$ is related to the quark condensate in the chiral limit by 
$\langle 0|\bar{q} q|0 \rangle = -F^2 B$.

The $(8_L,1_R)$ piece of the nonleptonic weak interactions is
represented by the well-known Cronin Lagrangian
\cite{Cronin},
\beq
F^2 \; \langle \Xi u_\mu u^\mu \rangle~, \qquad
\Xi =  G_8 F^2 \Delta + {\rm h.c.} \,  .
\eeq
At lowest order, the parameter $G_8$ can be determined \cite{PGR86}
from $K \ra 2 \pi$ decays to be 
$|G_8| \simeq 9 \times 10^{-6} {\rm GeV}^{-2} \simeq 5 (G_F / \sqrt{2}) 
|V_{ud} V_{us}|$.
 
Now also the electromagnetic interaction has to be included. Apart from the 
necessary modification in (\ref{photcoupl}), we have to add a kinetic term for
the photon field,
\beq
- \frac{1}{4} F_{\mu\nu} F^{\mu\nu}~, \qquad 
F_{\mu \nu} = \partial_{\mu} A_{\nu} - \partial_{\nu} A_{\mu}~,
\eeq
and a strangeness-conserving term of $\cO (e^2 p^0)$ \cite{EGPR89},
\begin{equation} \label{defZ}
e^2 F^4 Z \langle \Q_L \Q_R\rangle ~.
\eeq
The numerical value of the parameter $Z$ can be determined from 
the mass difference of charged and neutral pions. The relation 
$M_{\pi^{\pm}}^2 - M_{\pi^0}^2 =  2 e^2 Z F^2$  
implies $Z \simeq 0.8$.

Finally, we have to introduce a weak-electromagnetic term characterized by a 
coupling constant $g_{\rm ewk}$,
\begin{equation} \label{gewk}
e^2 F^4 \langle \Upsilon \Q_R\rangle~, \qquad
\Upsilon = g_{\rm ewk}G_8 F^2 \Delta + {\rm h.c.}  \, .
\eeq
Note that to lowest order only a single (linear independent) term of this type
can be constructed once the relation
\beq \label{relation1}
\Q_L \Delta = \Delta \Q_L = - \frac{1}{3} \Delta 
\eeq
is taken into account. This term is the lowest-order chiral
realization of electromagnetic penguins \cite{BW84,GRW}
and transforms as $(8_L,8_R)$ under $G$ 
when the $Q_R$ spurion field is ``frozen'' to 
the constant value (\ref{Qem}). By chiral
dimensional analysis we expect the coupling constant $g_{\rm ewk}$ to 
be of $\cO(1)$. A recent estimate in Ref.~\cite{CDG1} corresponds in
fact to $g_{\rm ewk}= - 1.0 \pm 0.3$ (see also Ref.~\cite{emplit}).

Summing up all these contributions, our lowest-order effective   
Lagrangian assumes the form
\beqa \label{L2}
\cL_2 &=& 
\frac{F^2}{4} \; \langle u_\mu u^\mu + \chi_+\rangle +
F^2 \; \langle \Xi u_\mu u^\mu \rangle  \no \\
&& \mbox{} - \frac{1}{4} F_{\mu\nu} F^{\mu\nu} +
e^2 F^4 Z \langle \Q_L \Q_R\rangle +
e^2 F^4 \langle \Upsilon \Q_R\rangle~.
\eeqa
Using (\ref{CPStransf}), one easily verifies that  
(\ref{L2}) is CPS invariant.

\section{One-loop divergences}
\label{sec:divs} 
\renewcommand{\theequation}{\arabic{section}.\arabic{equation}}
\setcounter{equation}{0}
For the construction of the one-loop functional, we first add a gauge-breaking
term (we are using the Feynman gauge) and external sources to
(\ref{L2}):
\beq
\cL_2 \ra \cL_2 - \frac{1}{2}(\partial_\mu A^\mu)^2 -
J_\mu A^\mu ~. \label{Lsource}
\eeq
Then we expand the lowest-order action associated with (\ref{Lsource}) 
around the solutions $\vp_{\rm cl}$, $A^\mu_{\rm cl}$ of the classical 
equations of motion.
In the standard ``gauge'' $u_R(\vp_{\rm cl}) = u_L(\vp_{\rm cl})^\dg =:
u(\vp_{\rm cl})$, a convenient choice of the pseudoscalar fluctuation
variables $\xi_i$ ($i = 1,\ldots,8$) is given by
\beq
u_R = u_{\rm cl} e^{i \xi_i \lambda_i/2F}~, \qquad
u_L = u^{\dg}_{\rm cl} e^{-i \xi_i \lambda_i/2F}~, \qquad
\xi_i(\vp_{\rm cl}) = 0~,
\eeq
with the Gell-Mann matrices $\lambda_i$ ($i = 1,\ldots,8$).
The photon field is decomposed as
\beq
A^\mu = A^\mu_{\rm cl} + \ve^\mu
\end{equation}
with a fluctuation field $\ve^\mu$.
In the following formulas, we shall drop the subscript ``${\rm cl}$''
for simplicity.
The classical equations of motion take the form
\beqa \label{EOM}
\nabla_\mu u^\mu &=& \frac{i}{2} \left(\chi_- -
\frac{1}{3} \langle \chi_-\rangle \right) + 
2 ie^2 F^2 Z [\Q_R, \Q_L]  \nn
&& + i [u_\mu u^\mu , \Xi ] - 2 ( \nabla_\mu \{ u^\mu , \Xi \} - \frac{1}{3}
\langle \nabla_\mu \{u^\mu , \Xi \} \rangle ) \nn
&& + 2 i e^2 F^2 [\Q_R, \Upsilon] ~,
\eeqa
\beqa \label{Maxwell}
\Box A_\mu &=& J_\mu + \frac{e F^2}{2} \langle u_\mu
(\Q_R - \Q_L)\rangle + e F^2 \langle \Xi \{\Q_R-\Q_L, u_\mu \} \rangle ~,  
\eeqa
where
\beqa
\nabla_\mu &=& \partial_\mu + [\Gamma_\mu,\quad]~, \no \\
\Gamma_\mu &=& \frac{1}{2} [u^\dg (\partial_\mu - i r_\mu)u
+ u(\partial_\mu - i l_\mu) u^\dg]~.\label{conn}
\eeqa
The solutions of (\ref{EOM}) and (\ref{Maxwell}) are uniquely 
determined functionals of the external sources 
$v_\mu$, $a_\mu$, $\chi$, $J_\mu$. (Note that the usual Feynman 
boundary conditions are always implicitly understood.)

Expanding (\ref{Lsource}) up to terms quadratic in the fields $\xi_i$,
$\ve_\mu$,  we obtain the second-order fluctuation Lagrangian $\cL^{(2)}$.
The one-loop functional $W_{L = 1}$ is then given by the Gaussian functional
integral
\beqa
\label{Gauss}
e^{iW_{L = 1}} =
\int [d\xi_i d \ve_\mu] ~
e^{i \int d^d x \cL^{(2)}}~.
\eeqa
In our case, $\cL^{(2)}$ reads
\beqa \label{Lfluc1}
\cL^{(2)} &=& \frac{F^2}{4} \langle \nabla_\mu \xi \nabla^\mu \xi
+ \frac{1}{2} u_\mu \xi u^\mu \xi - \frac{1}{2} (u_\mu u^\mu +\chi_+) \xi^2
\rangle \no \\
&& \mbox{} + e^2 F^4 Z \langle \xi \Q_L \xi \Q_R - \frac{1}{2} \xi^2 
\{\Q_L,\Q_R\} \rangle \no \\
&& \mbox{} + \frac{F^2}{4} \langle 4 \Xi (\nabla_\mu \xi \nabla^\mu \xi
- \frac{1}{4} \{ u_\mu ,\{u^\mu, \xi^2\}\} 
+ \frac{1}{4} \{u_\mu,\xi\} \{u^\mu,\xi\})
- 2 i [\xi,\Xi] \{\nabla_\mu \xi, u^\mu \}  \rangle \no \\
&& \mbox{} + e^2 F^4 \langle \Upsilon (\xi \Q_R \xi - \frac{1}{2} 
\{\xi^2,\Q_R\}) \rangle \no \\
&& \mbox{} + \frac{1}{2} \ve_\mu \Box \ve^\mu + \frac{e^2 F^2}{4}
\langle (\Q_R - \Q_L)^2\rangle \ve_\mu \ve^\mu 
+ e^2 F^2 \langle \Xi (\Q_R - \Q_L)^2 \rangle \ve_\mu \ve^\mu \no \\
&& \mbox{} - \frac{ie F^2}{4} \langle [u_\mu, \Q_R + \Q_L] \xi 
\rangle \ve^\mu  + \frac{e F^2}{2} \langle (\Q_R - \Q_L) 
\nabla_\mu \xi\rangle \ve^\mu \no \\
&& \mbox{} - \frac{i e F^2}{2} \langle \Xi \{ [ \Q_R + \Q_L, \xi],
 u_\mu \} \rangle \ve^\mu - \frac{i e F^2}{2} \langle [\xi, \Xi] 
\{ \Q_R - \Q_L, u_\mu\} \rangle \ve^\mu \no \\
&& \mbox{} + e F^2 \langle \Xi \{ \Q_R - \Q_L, \nabla_\mu \xi \} 
\rangle \ve^\mu ~,
\eeqa
where
\beq
\xi = \xi_i \lambda_i /F ~.
\eeq
In the next step, we perform the field transformation
\beq
\xi \ra \xi - \{ \Xi, \xi \} + \frac{2}{3} \langle \Xi \xi \rangle {\bf 1} ~.
\eeq
Because of $\langle \Delta \rangle = 0$, we do not pick up an additional
contribution from the Jacobi determinant and
the fluctuation Lagrangian (\ref{Lfluc1}) assumes the form
\beq \label{Lfluc2}
\cL^{(2)} = - \frac{1}{2} \xi_i (d_\mu d^\mu + \sigma)_{ij} \xi_j
+ \frac{1}{2} \ve_\mu (\Box  + \kappa ) \ve^\mu 
+ \ve_\mu a^\mu_i \xi_i + \ve_\mu b_i d^\mu_{ij} \xi_j ~,
\eeq
where
\beq \label{covdev}
d^{\mu}_{ij} = \delta_{ij} \partial^{\mu} + \gamma^{\mu}_{ij}~. 
\eeq
The explicit expressions for $\gamma^{\mu}_{ij}$, $\sigma_{ij}$, 
$\kappa$, $a^\mu_i $, $b_i$ are given in the Appendix.

The divergent part of the one-loop functional,
\begin{equation} \label{divfunc}
W_{L = 1}^{\rm div} = \int d^d x \cL^{\rm div}_{L=1}~,
\eeq
is determined by
\beqa 
\cL^{\rm div}_{L=1} &=& 
 - \dfrac{1}{{(4\pi)}^2 (d-4)}  
[ {\rm tr} (\frac{1}{12} \gamma_{\mu \nu}  \gamma^{\mu \nu} 
+ \frac{1}{2} \sigma^2 ) 
- a_i^{\mu} a_{\mu i} + a_i^{\mu} (d_\mu b)_i  \nn
&& \qquad \qquad  \qquad  \quad + \frac{1}{2} (b_i b_i)^2
- b_i \sigma_{ij} b_j - \kappa b_i b_i  + 2 \kappa^2  ] ~,
\label{Ldiv}
\eeqa
where
\beq
\gamma_{\mu \nu} = \partial_\mu \gamma_\nu - \partial_\nu \gamma_\mu
+ [\gamma_\mu, \gamma_\nu] ~.
\eeq
This formula can easily be derived from the well-known second Seeley-deWitt
coefficient for bosonic systems \cite{Ball}.

\section{The chiral Lagrangian at next-to-leading order}
\label{sec:NLO}
\renewcommand{\theequation}{\arabic{section}.\arabic{equation}}
\setcounter{equation}{0}
We are now in the position to construct the most general local
action at next-to-leading order which will also renormalize the 
one-loop divergences discussed in the previous section.

The strong part of the local action of $\cO(p^4)$ is, of course, generated by
the well-known Gasser-Leutwyler Lagrangian \cite{GL85} associated
with the low-energy constants $L_1, \dots, L_{12}$.  
In the presence of virtual photons, the structure of the operators
given in \cite{GL85} remains unchanged. The only necessary
modification is the inclusion of the dynamical photon field in the
generalized ``sources" $\ell_\mu$ and $r_\mu$ (see (\ref{sources})).
The divergences corresponding to  the strong sector of (\ref{Ldiv}) are
absorbed by the divergent parts of the $L_i$ \cite{GL85}. 
In the relevant case of chiral $SU(3)$, the strong terms generated by
(\ref{Ldiv}) can be written immediately as a linear combination of the
$\cO(p^4)$ operators of the Gasser-Leutwyler basis without 
using the equations of motion (\ref{EOM}) or (\ref{Maxwell}).
Consequently, no additional (weak-)electromagnetic terms are induced at
this point.

The strangeness-conserving terms of  $\cO(e^2 p^2)$ have been constructed by
Urech \cite{Urech}. His list of electromagnetic counterterms is
associated with the coupling constants $K_1, \dots, K_{14}$. In this
case, (\ref{Ldiv}) leads to that canonical basis only after the use of
the equation of motion (\ref{EOM}). In this way,
also some divergent weak-electromagnetic contributions of $\cO(G_8 e^2
p^2)$ are generated.

For the octet part of the nonleptonic weak Lagrangian of $\cO(G_F p^4)$ 
\cite{KMW90} we refer to the standard form of Ecker, Kambor and 
Wyler \cite{EKW93} with couplings $N_1, \dots, N_{37}$. Again, because of
the mismatch between (\ref{Ldiv}) and the standard basis, the equation of
motion has to be used and the (purely) electromagnetic piece in
(\ref{EOM}) induces divergent terms of $\cO(G_8 e^2 p^2)$.

Finally, we have to construct the most general
weak-electromagnetic Lagrangian of $\cO(G_8 e^2 p^2)$. Some parts of
this Lagrangian have appeared before in the literature
\cite{EdR89,CDG2,AH}. 
The complete minimal Lagrangian of $\cO(G_8 e^2 p^2)$
takes the form
\begin{equation} \label{lagewk}
\cL_{G_8 e^2 p^2} =  G_8 e^2 F^4 \sum_{i=1}^{32} Z_i Q_i + {\rm h.c.} ~ ,
\eeq 
with operators $Q_i$ of $\cO(p^2)$ and dimensionless coupling 
constants $Z_i$.
A linear independent set of  operators  is given by
\beqa \label{list}
Q_1 &=& \langle \Delta \{ \Q_R, \chi_+ \} \rangle~, \nn
Q_2 &=& \langle \Delta \Q_R \rangle \langle \chi_+ \rangle~, \nn
Q_3 &=& \langle \Delta \Q_R \rangle \langle \chi_+ \Q_R \rangle~, \nn
Q_4 &=& \langle \Delta \chi_+ \rangle \langle \Q_L \Q_R \rangle~, \nn
Q_5 &=& \langle \Delta u_\mu u^\mu \rangle~, \nn
Q_6 &=& \langle \Delta \{\Q_R,u_\mu u^\mu\} \rangle~, \nn
Q_7 &=& \langle \Delta u_\mu u^\mu \rangle \langle \Q_L \Q_R \rangle~, \nn
Q_8 &=& \langle \Delta u_\mu \rangle \langle \Q_L u^\mu \rangle~, \nn
Q_9 &=& \langle \Delta u_\mu \rangle \langle \Q_R u^\mu \rangle~, \nn
Q_{10} &=& \langle \Delta u_\mu \rangle \langle \{\Q_L,\Q_R\}u^\mu\rangle~, \nn
Q_{11} &=& \langle \Delta \{\Q_R,u_\mu\} \rangle \langle \Q_L u^\mu \rangle~,\nn
Q_{12} &=& \langle \Delta \{\Q_R,u_\mu\} \rangle \langle \Q_R u^\mu \rangle~,\nn
Q_{13} &=& \langle \Delta \Q_R \rangle \langle u_\mu u^\mu \rangle~, \nn
Q_{14} &=& \langle \Delta \Q_R \rangle \langle u_\mu u^\mu \Q_R \rangle~, \nn
Q_{15} &=& \langle \Delta \Q_R \rangle \langle u_\mu u^\mu (\Q_L-\Q_R) \rangle~,\nn
Q_{16} &=& \langle \Delta \chi_+ \rangle~, \nn
Q_{17} &=& \frac{2}{3} \langle \Delta \chi_+ \rangle   
 + \langle \Delta \{ \Q_R, \chi_+ \} \rangle 
 + \langle \Delta [ \Q_R, \chi_- ] \rangle~, \nn
Q_{18} &=& \langle \Delta \{ \Q_R, \chi_+ \} \rangle 
 -\frac{1}{3}  \langle \Delta [ \Q_R, \chi_- ] \rangle 
 +\langle \Delta (\chi_-\Q_L\Q_R-\Q_R\Q_L\chi_-)\rangle \nn
&& \mbox{} -\frac{4}{3} \langle \Delta \Q_R \rangle \langle \chi_+ \rangle 
 - \langle \Delta \Q_R \rangle \langle \chi_+ \Q_R \rangle 
 +\langle \Delta \chi_+ \rangle \langle \Q_L \Q_R \rangle~, \nn
Q_{19} &=&  \langle \Delta \Q_R \rangle \langle \chi_+ (\Q_L -\Q_R) \rangle~, \nn
Q_{20} &=& i \langle (\wh{\nabla}_\mu \Delta) [\Q_L,u^\mu]\rangle~, \nn
Q_{21} &=& i \langle (\wh{\nabla}_\mu \Delta) [\Q_R,u^\mu]\rangle~, \nn
Q_{22} &=& i \langle (\wh{\nabla}_\mu \Delta) (\Q_L u^\mu \Q_R
             - \Q_R u^\mu \Q_L) \rangle~, \nn
Q_{23} &=& i \langle (\wh{\nabla}_\mu \Delta) (u^\mu \Q_L \Q_R
             - \Q_R \Q_L u^\mu) \rangle~, \nn
Q_{24} &=& i \langle \Delta (u_\mu (\wh{\nabla}^\mu \Q_L) \Q_R
                            - \Q_R (\wh{\nabla}^\mu \Q_L) u_\mu) \rangle~, \nn
Q_{25} &=& i \langle (\wh{\nabla}_\mu \Delta) (u^\mu \Q_R \Q_L
             - \Q_L \Q_R u^\mu) \rangle~, \nn
Q_{26} &=& i \langle \Delta (u_\mu \Q_R (\wh{\nabla}^\mu \Q_R) 
                            - (\wh{\nabla}^\mu \Q_R) \Q_R u_\mu) \rangle~, \nn
Q_{27} &=& i \langle \Delta (\Q_R u_\mu (\wh{\nabla}^\mu \Q_R)
                            - (\wh{\nabla}^\mu \Q_R) u_\mu \Q_R) \rangle~, \nn
Q_{28} &=& \langle (\wh{\nabla}_\mu \Delta) (\wh{\nabla}^\mu \Q_L) \rangle~,\nn
Q_{29} &=& \langle (\wh{\nabla}_\mu \Delta) (\wh{\nabla}^\mu \Q_R) \rangle~,\nn
Q_{30} &=&\langle\Delta (\wh{\nabla}_\mu \Q_R)(\wh{\nabla}^\mu \Q_R)\rangle~,\nn
Q_{31} &=& \langle (\wh{\nabla}^\mu \Delta) \{\wh{\nabla}_\mu \Q_L,
\Q_R\} \rangle~, \nn
Q_{32} &=& \langle (\wh{\nabla}^\mu \Delta) \{\wh{\nabla}_\mu \Q_R,
\Q_L\} \rangle~, 
\eeqa
where
\beqa \label{covder1}
\wh \nabla_\mu \Delta &=& \nabla_\mu \Delta 
+ \frac{i}{2} [u_\mu,\Delta] =
u (D_\mu \lambda) u^\dg ~, \no \\
\wh \nabla_\mu \Q_L &=& \nabla_\mu \Q_L 
+ \frac{i}{2} [u_\mu,\Q_L] =
u (D_\mu Q_L) u^\dg ~, \no \\
\wh \nabla_\mu \Q_R &=& \nabla_\mu \Q_R 
- \frac{i}{2} [u_\mu,\Q_R] =
u^\dg (D_\mu Q_R) u ~,
\eeqa
with
\beqa \label{covder2}
D_\mu \lambda &=& \partial_\mu \lambda 
- i[l_\mu, \lambda]~, \nn
D_\mu Q_L &=& \partial_\mu Q_L 
- i[l_\mu,Q_L]~, \nn
D_\mu Q_R &=& \partial_\mu Q_R - i[r_\mu,Q_R]~.
\eeqa

For the construction of the list of local terms (\ref{list}) we have used CPS
invariance, the  relations (\ref{relation1}) and
\beq
\Q_{L,R}^2 = \frac{2}{9} {\bf 1} + \frac{1}{3} \Q_{L,R}~,
\eeq 
the Cayley-Hamilton formula, partial integration and the equations of motion
(\ref{EOM}). 

If the spurion fields $Q_{L,R}$ and $\lambda$ 
are fixed to the constant values in (\ref{Qem})
and (\ref{lambda}), respectively, then $\cL_{G_8 e^2 p^2}$ 
transforms under $G$ as
\beq
(8_L, 1_R) + (8_L, 8_R) + (27_L, 1_R) + (27_L, 8_R) + (8_L, 27_R)~.
\eeq
This structure is richer than the one of the 
$\cO(G_8)$ terms in $\cL_2$ and  also of the weak
four-fermion effective Hamiltonian \cite{OPE}.
The last two pieces, in particular, which are  
responsible for $\Delta I=5/2$ transitions,
have no analog in the effective Hamiltonian of dimension six. 

The operator $Q_{16}$ does not contribute to on-shell matrix elements
\cite{CPS,Crewther,Leurer,KMW90}. 
The terms $Q_{17}$, $Q_{18}$, $Q_{19}$ vanish for electrically
neutral (pseudo)scalar sources,
\beq
[\chi,Q] = 0~,
\eeq
which is, of course, the case for all realistic physical processes. Also the 
operators $Q_{20}, \dots Q_{32}$ are irrelevant for practical purposes.
Because of
(\ref{covder1}) and (\ref{covder2}), they
contribute only in the presence of non-vanishing external (axial-)vector
sources.

The coupling constants $Z_1, \dots , Z_{12}$ appear in the amplitudes 
of $K \to 2 \pi$ decays. The operators $Q_{13}$ and $Q_{14}$ do not
contribute to $K \to 2 \pi$ but they enter for $K \to 3 \pi$. $Q_{15}$
involves at least five pseudoscalar fields and is therefore irrelevant
for $K$ decays.
A few linear combinations of the operators in (\ref{list}) were
already given some time ago by de Rafael \cite{EdR89}. His
list was restricted to terms contributing to $K \to 2 \pi$, neglecting
contributions $\sim M_\pi^2$ and those renormalizing $G_8$.
A more recent extension of de Rafael's list can be found in 
Ref.~\cite{CDG2}. However, their Lagrangian is still incomplete even for 
the $K \to 2 \pi$ amplitudes, as we shall discuss in the following
section. There is in addition an obvious misprint in the 
operator multiplied by $s_6$ in \cite{CDG2}, which would be in conflict 
with chiral symmetry. Some of the operators in (\ref{list}) have also 
appeared in attempts \cite{AH} to bosonize the $\Delta S=1$
four-fermion effective Hamiltonian.

The low-energy couplings $Z_i$  are in general divergent. 
They absorb the divergences of the 
one-loop graphs via the renormalization 
\beqa
Z_i &=& Z_i^r(\mu) + z_i \Lambda(\mu) ~, \quad i=1,\ldots,32 ~, \no \\
\Lambda(\mu) &=& \frac{\mu^{d-4}}{(4\pi)^2} \left\{ \frac{1}{d-4} -
\frac{1}{2} [\ln (4\pi) + \Gamma'(1) + 1]\right\}, \label{renorm}
\eeqa
in the dimensional regularization scheme.   
The  coefficients $z_1, \ldots z_{32}$  are 
determined in such a way that the divergences generated by  (\ref{Ldiv}) are 
cancelled:
\beq \label{zs}
\ba{llll}
z_1 =  -\dfrac{17}{12} -3 Z +\dfrac{3}{2} g_{\rm ewk}~, &
z_2 = 1 + \dfrac{16}{3} Z + g_{\rm ewk}~, & 
z_3 = \dfrac{3}{4} + 7 Z~, &
z_4 =  -\dfrac{3}{4} - 7 Z~, 
\\ [10pt]  
z_5 = - 2 ~, & 
z_6 =  \dfrac{7}{2} + 5 Z + \dfrac{3}{2} g_{\rm ewk}~, &
z_7 = \dfrac{3}{2} + 5 Z~, &
z_8 = -\dfrac{1}{2}~, 
\\ [10pt]
z_9 = -\dfrac{11}{6} + \dfrac{4}{3} Z + 2 g_{\rm ewk}~, &
z_{10} = -\dfrac{3}{2} - Z~, &
z_{11} = -\dfrac{3}{2} - 2 Z~, &
z_{12} = \dfrac{3}{2}~, 
\\ [10pt]
z_{13} = -\dfrac{35}{12} - 3 Z + g_{\rm ewk}~, &
z_{14} = 3 + 15 Z~, &
z_{15} = \dfrac{3}{2} + 15 Z~, &
z_{16} = -\dfrac{4}{9} - \dfrac{4}{3} Z~, 
\\ [10pt]
z_{17} = \dfrac{2}{3} + 2 Z~, &
z_{18} = \dfrac{3}{4} + 3 Z~, &
z_{19} = 4 Z~, &
z_{20} = -\dfrac{1}{2}~,
\\ [10pt]
z_{21} = \dfrac{1}{6}~, &
z_{22} = 3 + 6 Z~, &
z_{23} = -3 - 9 Z~, &
z_{24} = 0~,
\\ [10pt]
z_{25} =  - 3 Z~, &
z_{26} = -1~, &
z_{27} = 0~, &
z_{28} = -\dfrac{1}{2}~,
\\ [10pt]
z_{29} = -\dfrac{1}{2}~, &
z_{30} =  0~, &
z_{31} = \dfrac{3}{2} + 6 Z~, &
z_{32} = \dfrac{3}{2} + 6 Z~.
\ea
\eeq

As already discussed above, the values in this list depend on our
conventions for the basis systems in the strong, electromagnetic and weak
parts of the next-to-leading order Lagrangian. The $z_i$ given in
(\ref{zs}) have to be used together with the divergent parts of
the coupling constants $L_i$ \cite{GL85}, $K_i$ \cite{Urech} and $N_i$
\cite{EKW93}, respectively. The divergences involving the electroweak
penguin coupling $g_{\rm ewk}$ are independent of this choice of basis
and they agree with a recent calculation of Cirigliano
and Golowich \cite{CG00}. 
Note that  $g_{\rm ewk}$ appears only in the couplings of 
$(8_L,8_R)$ operators. This is because  
the lowest-order term proportional to $g_{\rm ewk}$
is already of $\cO(G_8e^2)$. Therefore,  
the $\cO(G_8e^2p^2)$ terms proportional 
to  $g_{\rm ewk}$ arise from the 
product of the lowest-order $(8_L,8_R)$ 
weak operator times the $\cO(p^2)$ invariant 
part of the strong Lagrangian.

The renormalized low-energy constants $Z_i^r(\mu)$ are in general 
scale dependent. The coefficients $z_i$ govern this scale
dependence through the renormalization group equations
\begin{equation} 
\mu \displaystyle\frac{d Z_i^r(\mu)}{d\mu} =
- \displaystyle\frac{z_i}{(4\pi)^2} ~.
\end{equation} 
By construction, the complete generating functional at 
next-to-leading order is then scale independent.

\section{$K \ra \pi \pi$}
\label{sec:decays}
\renewcommand{\theequation}{\arabic{section}.\arabic{equation}}
\setcounter{equation}{0}

In the modern framework of chiral perturbation theory, electromagnetic 
corrections for $K \to \pi \pi$ decays to $\cO(G_8 e^2 p^2)$ were 
discussed by de Rafael \cite{EdR89} and have been treated in more
detail by Cirigliano, Donoghue and Golowich \cite{CDG1,CDG2,CDG3}. 
Together with corrections of $\cO(G_8 (m_u-m_d) p^2)$ \cite{WM99}, 
the complete isospin-breaking effects of next-to-leading order have 
obvious phenomenological implications, from the $\Delta I=1/2$ rule 
to CP violation \cite{EMNP1}.

In this section, we present the tree-level contributions to the
$K \to \pi\pi$ amplitudes from the Lagrangian (\ref{lagewk}). We
compare those amplitudes and in particular their divergent parts
with the results of Ref.~\cite{CDG2}. Using our own one-loop
calculation of isospin-breaking corrections \cite{EIMNP2} and the
heat-kernel results (\ref{zs}), we find that the complete amplitudes
of $\cO(G_8 e^2 p^2)$ are indeed finite. We demonstrate the cancellation
of divergences explicitly for the subset of amplitudes proportional to
the electromagnetic penguin coupling $g_{\rm ewk}$ defined in (\ref{gewk}).

From the Lagrangian (\ref{lagewk}) of $\cO(G_8 e^2 p^2)$, we obtain the
following amplitudes in units of $C_{\rm ewk}:= i G_8 e^2 F$:
\begin{eqnarray} \label{KppZ}
A(K^0 \to \pi^+ \pi^-) &=& C_{\rm ewk}\sqrt{2}\left[(M_K^2-M_\pi^2)
( 2 Z_1 + 4 Z_2 - 4/3 Z_3 + 4 Z_4 - Z_5 - 1/3 Z_6  \right. \nn
& & \left. - 2/3 Z_7) + M_\pi^2 (6 Z_1 + 6 Z_2 - Z_6)\right] ~,\nn
A(K^0 \to \pi^0 \pi^0) &=& C_{\rm ewk}\sqrt{2}(M_K^2-M_\pi^2)
(  - Z_5 + 2/3 Z_6 - 2/3 Z_7 + Z_8 +  Z_9 + 2/3 Z_{10}  \nn
& &  - 2/3 Z_{11} - 2/3 Z_{12} )  ~,\nn
A(K^+ \to \pi^+ \pi^0) &=& C_{\rm ewk} \left[(M_K^2-M_\pi^2)
( 2 Z_1 + 4 Z_2 - 4/3 Z_3 - Z_6 - Z_8 - Z_9 - 2/3  Z_{10} \right. \nn
& & \left. - 4/3 Z_{11} - 4/3 Z_{12} ) + M_\pi^2 (6 Z_1 + 6 Z_2
-Z_6) \right] ~.
\end{eqnarray} 
These amplitudes agree with Ref.~\cite{CDG2} for $Z_3=3 Z_2, Z_{10}=
Z_{11}=0$. In addition, the coefficients $s_8, s_9$ in Eq.~(35) of
\cite{CDG2} should be multiplied by 2/3.

In the $SU(3)$ limit for the mass matrix (\ref{Mquark}), 
the amplitudes (\ref{KppZ}) satisfy the relations
\begin{eqnarray} 
A(K^0 \to \pi^+ \pi^-)_{SU(3)} &=& 
\sqrt{2} A(K^+ \to \pi^+ \pi^0)_{SU(3)} ~,\nn
A(K^0 \to \pi^0 \pi^0)_{SU(3)} &=& 0 ~, \label{su3}
\end{eqnarray} 
in accordance with a general theorem on $K\to \pi\pi$ transitions in 
the presence of electromagnetism \cite{EIMNP2}.

The divergent parts of the $Z_i$ in (\ref{zs}) give rise to the 
following divergent tree-level amplitudes, with 
$\Lambda(\mu)$ and $Z$ defined in (\ref{renorm}) and (\ref{defZ}),
respectively:
\begin{eqnarray} \label{Kppdiv}
A(K^0 \to \pi^+ \pi^-)_{\rm div} &=& C_{\rm ewk}\sqrt{2}\Lambda(\mu)\left[
M_K^2 ( - 3 - 27 Z + 13/2 g_{\rm ewk}) \right. \nn
& & \left. + M_\pi^2  ( - 3 + 36 Z + 7 g_{\rm ewk})\right]~, \nn
A(K^0 \to \pi^0 \pi^0)_{\rm div} &=& C_{\rm ewk}\sqrt{2}\Lambda(\mu)
(M_K^2 - M_\pi^2) ( 2 Z + 3 g_{\rm ewk} ) ~,\\
A(K^+ \to \pi^+ \pi^0)_{\rm div} &=& C_{\rm ewk}\Lambda(\mu)\left[
M_K^2 ( 3 Z + 7/2 g_{\rm ewk} )
+ M_\pi^2 ( - 6 + 6 Z + 10 g_{\rm ewk} )\right]~.\no
\end{eqnarray}

The (ultraviolet) divergences in (\ref{Kppdiv}) arise from three 
different sources:
\begin{itemize} 
\item Photon loops proportional to $G_8 e^2$;
\item Loops involving the electromagnetic coupling 
(\ref{defZ}) proportional to $G_8 e^2 Z$;
\item Loops involving the coupling 
(\ref{gewk}) proportional to $G_8 e^2 g_{\rm ewk}$.
\end{itemize}
Strong and electromagnetic wave function renormalization \cite{NR95}
is included in all three
categories. 

We have performed a complete calculation of $K \to \pi \pi$ 
amplitudes to $\cO(G_8 e^2 p^2)$ and $\cO(G_8 (m_u-m_d) p^2)$
\cite{EIMNP2}. For $m_u=m_d$, we find that the explicit loop 
divergences are exactly cancelled by the divergent tree-level amplitudes
(\ref{Kppdiv}). We exhibit those cancellations in detail for
the divergences proportional to $g_{\rm ewk}$. 
Divergences arise both in loops with an electromagnetic penguin vertex
shown in Fig.~\ref{loopsewk} and from (strong) wave function
renormalization of tree diagrams from the Lagrangian (\ref{gewk}).

\begin{figure}
\centerline{\epsfig{file=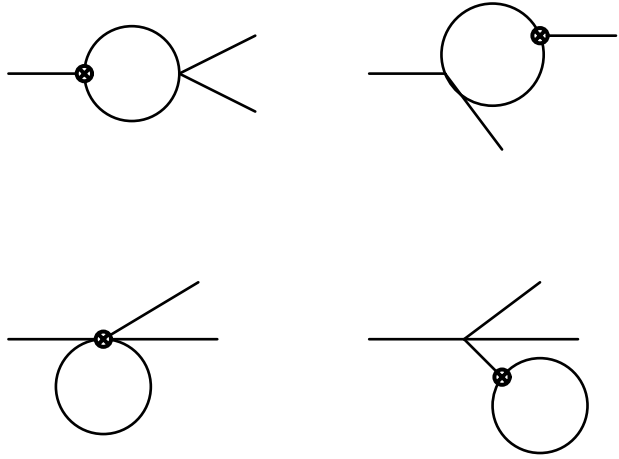,height=6cm}}
\caption{Loop diagrams for $K \to \pi\pi$ involving the
electromagnetic penguin coupling $g_{\rm ewk}$. The associated vertex
from the Lagrangian (\protect\ref{gewk}) is denoted by a crossed 
circle. Normal vertices are from the
lowest-order strong Lagrangian (\protect\ref{L2strong}).}
\label{loopsewk}
\end{figure}

In the exponential parametrization, the divergences due to the
diagrams of Fig.~\ref{loopsewk} take the form
\begin{eqnarray} \label{irrloops}
A(K^0 \to \pi^+ \pi^-)_{\rm loops} &=& - \displaystyle\frac{\sqrt{2}}
{2} C_{\rm ewk} g_{\rm ewk} \Lambda(\mu)( 7 M_K^2 + 8 M_\pi^2) ~,\nn
A(K^0 \to \pi^0 \pi^0)_{\rm loops} &=& -3 \sqrt{2} C_{\rm ewk}
g_{\rm ewk} \Lambda(\mu) (M_K^2 - M_\pi^2) ~,\nn
A(K^+ \to \pi^+ \pi^0)_{\rm loops} &=& - C_{\rm ewk} g_{\rm ewk}
\Lambda(\mu) (M_K^2/2 +  7 M_\pi^2)~.
\end{eqnarray}
Wave function renormalization (again in exponential parametrization)
leads to
\begin{eqnarray} \label{wfr}
A(K^0 \to \pi^+ \pi^-)_{\rm wfr} &=& - 3 \sqrt{2} C_{\rm ewk} 
g_{\rm ewk} \Lambda(\mu)( M_K^2 + M_\pi^2) ~,\nn
A(K^0 \to \pi^0 \pi^0)_{\rm wfr} &=& 0 ~,\nn
A(K^+ \to \pi^+ \pi^0)_{\rm wfr} &=& - 3 C_{\rm ewk} g_{\rm ewk}
\Lambda(\mu) (M_K^2 +  M_\pi^2)~.
\end{eqnarray}
The sum of (\ref{irrloops}) and (\ref{wfr}) is parametrization 
independent and it is exactly cancelled by the terms in (\ref{Kppdiv}) 
proportional to $g_{\rm ewk}$.

We have exhibited (part of) the loop divergences explicitly also
because we do not completely agree with the results of 
Ref.~\cite{CDG2}. Although the divergences due to photon loops
are identical, we obtain different results for some of the other
divergences\footnote{V. Cirigliano has informed us that they now agree
with the divergences (\ref{Kppdiv}); see forthcoming erratum for 
Ref.~\cite{CDG2}.}. 
Only for the channel $K^0 \to \pi^+ \pi^-$, there is
complete agreement for all three types of divergences.

The complete amplitudes of $\cO(G_8 e^2 p^2)$ and 
$\cO(G_8 (m_u-m_d) p^2)$ together with a phenomenological 
analysis will be presented elsewhere \cite{EIMNP2}.

\section{Conclusions}
\label{sec:Conclusions}
\renewcommand{\theequation}{\arabic{section}.\arabic{equation}}
\setcounter{equation}{0}

We have supplied the missing ingredients for a complete analysis 
at next-to-leading order of the combined strong, nonleptonic weak
and electromagnetic interactions of mesons. The main results are:
\begin{enumerate} 
\item[i.] The complete and minimal Lagrangian (\ref{lagewk})
of $\cO(G_8 e^2 p^2)$ contains 32 operators $Q_i$ and
associated dimensionless coupling constants $Z_i$. Of these 32
operators, only 14 are of immediate 
phenomenological relevance. We have ordered the terms in a
way most suitable for applications: the first 12 operators 
contribute to $K \to 2 \pi$ decays whereas the remaining two 
enter in $K \to 3 \pi$ amplitudes.
\item[ii.] The one-loop divergence functional 
(\ref{divfunc}) determines the renormalization of the effective
theory. Together with the previously known divergences, the new 
terms (\ref{zs}) in the 
coupling constants $Z_i$ ensure that the complete amplitudes
for strong, nonleptonic weak and electromagnetic interactions of
mesons at next-to-leading order are finite. 
\end{enumerate} 

As a first application, we have presented the tree-level amplitudes
of $\cO(G_8 e^2 p^2)$ for $K \to \pi\pi$ decays. The associated
divergent parts cancel with the explicit one-loop divergences
\cite{EIMNP2} to yield finite and scale independent decay 
amplitudes. 

\vspace*{2cm}

\section*{Acknowledgements}

We thank J. Gasser and E. de Rafael for having started this project
with two of us (G.E., A.P.) some years ago. We are also grateful to
J. Bijnens, V. Cirigliano and J. Gasser for helpful correspondence.
H.N. acknowledges financial support from the University of Valencia 
through a ``Visiting Professorship'' and the IFIC Department of 
Theoretical Physics where part of this work was done for hospitality.

\medskip\medskip
\newcounter{zaehler}
\renewcommand{\thesection}{\Alph{zaehler}}
\renewcommand{\theequation}{\Alph{zaehler}.\arabic{equation}}
\setcounter{zaehler}{1}
\setcounter{equation}{0}

\section*{Appendix}
\label{appdx}

The quantities occurring in (\ref{Ldiv}) can be decomposed with
respect to (explicit\footnote{Note that $e$ also appears in the 
vielbein $u_\mu$ (\ref{photcoupl}) and in the connection 
$\Gamma_\mu$ (\ref{conn}) via (\ref{sources}).})
powers of $e$ and $G_8$ in the following way:
\beqa \label{decomposition}
\sigma_{ij} &=& \sigma_{ij}|_{e^0G_8^0} +  \sigma_{ij}|_{e^2G_8^0} +
                \sigma_{ij}|_{e^0G_8} +  \sigma_{ij}|_{e^2G_8}~, \nn
\gamma_\mu &=& \gamma_\mu|_{e^0G_8^0} + \gamma_\mu|_{e^0G_8}~, \nn
a_i^\mu &=& a_i^\mu|_{eG_8^0} + a_i^\mu|_{eG_8}~, \nn
b_i &=& b_i|_{eG_8^0} + b_i|_{eG_8}~, \nn
\kappa &=& \kappa|_{e^2G_8^0} + \kappa|_{e^2G_8}~.
\eeqa
The explicit expressions for the various terms are given by
\beqa 
\sigma_{ij}|_{e^0G_8^0} &=&  \frac{1}{8} \langle (u_\mu u^\mu + \chi_+) 
\{\lambda_i,\lambda_j\} \rangle
- \frac{1}{4} \langle u_\mu \lambda_i u^\mu \lambda_j \rangle ~, \\
\sigma_{ij}|_{e^2G_8^0} &=& e^2F^2Z
\langle \frac{1}{2}  \{\Q_R,\Q_L\}
\{ \lambda_i,\lambda_j\} 
 - \lambda_i \Q_R \lambda_j \Q_L -
\lambda_j \Q_R \lambda_i \Q_L \rangle ~, \\ 
\sigma_{ij}|_{e^0G_8} &=&
\frac{1}{4} \langle \{ \Xi, \lambda_i \} u_\mu \lambda_j u^\mu \rangle
+ \frac{1}{4} \langle \{ \Xi, \lambda_j \} u_\mu \lambda_i u^\mu \rangle
\nn
&& \mbox{}
-\frac{1}{8} \langle (u_\mu u^\mu + \chi_+) (\{ \lambda_i,\{ \Xi,\lambda_j \}\}
+ \{ \lambda_j,\{ \Xi,\lambda_i \}\}) \rangle
\nn
&& \mbox{}
+\frac{1}{6} \langle \Xi \lambda_i \rangle \langle \chi_+ \lambda_j \rangle
+\frac{1}{6} \langle \Xi \lambda_j \rangle \langle \chi_+ \lambda_i \rangle
\nn
&& \mbox{}
+\frac{1}{4} \langle \Xi \{u_\mu ,\{u^\mu,\{\lambda_i,\lambda_j\}\}\}\rangle
\nn
&& \mbox{}
-\frac{1}{4} \langle \Xi (\{u_\mu,\lambda_i\}\{u^\mu,\lambda_j\}
 + \{u_\mu,\lambda_j\}\{u^\mu,\lambda_i\}) \rangle
\nn
&& \mbox{}
+\frac{i}{4} \langle [u_\mu,\nabla^\mu\Xi]\{\lambda_i,\lambda_j\}\rangle
+\frac{i}{4} \langle [\nabla^\mu u_\mu,\Xi]\{\lambda_i,\lambda_j\}\rangle
\nn
&& \mbox{}
-\frac{1}{2} \langle \{\lambda_i,\lambda_j\}\nabla_\mu\nabla^\mu\Xi\rangle~,
\\
\sigma_{ij}|_{e^2G_8} &=& e^2 F^2 Z \langle \Xi (
\lambda_i \Q_R \lambda_j \Q_L + \lambda_j \Q_R \lambda_i \Q_L 
+ \Q_R \lambda_i \Q_L \lambda_j + \Q_R \lambda_j \Q_L \lambda_i \nn
&& \mbox{}
+ \lambda_i \Q_L \lambda_j \Q_R + \lambda_j \Q_L \lambda_i \Q_R 
+ \Q_L \lambda_i \Q_R \lambda_j + \Q_L \lambda_j \Q_R \lambda_i \nn
&& \mbox{}
- \lambda_i \{\Q_L,\Q_R\} \lambda_j - \lambda_j \{\Q_L,\Q_R\} \lambda_i 
-\frac{1}{2} \{ \{\Q_L,\Q_R\}, \{\lambda_i,\lambda_j\} \} ) \rangle \nn
&& \mbox{}
+ e^2 F^2 \langle \Upsilon (\frac{1}{2}\{\{\lambda_i,\lambda_j\},\Q_R\}-
\lambda_i \Q_R \lambda_j - \lambda_j \Q_R \lambda_i) \rangle ~,\\
\gamma^\mu_{ij}|_{e^0G_8^0} &=& -\frac{1}{2} \langle \Gamma^\mu
[\lambda_i,\lambda_j] \rangle ~,\\
\gamma^\mu_{ij}|_{e^0G_8} &=&
\frac{i}{4} \langle [\lambda_i,\Xi]\{u^\mu,\lambda_j\} \rangle -
\frac{i}{4} \langle [\lambda_j,\Xi]\{u^\mu,\lambda_i\} \rangle ~,\\
a_i^\mu|_{eG_8^0} &=&
-\frac{ieF}{4} \langle u^\mu [\Q_R+\Q_L,\lambda_i] \rangle ~,\\
a_i^\mu|_{eG_8} &=&
\frac{ieF}{4} \langle\Xi(u^\mu\lambda_i\Q_R - \Q_R\lambda_iu^\mu)\rangle
+\frac{3ieF}{4} \langle\Xi(u^\mu\lambda_i\Q_L - \Q_L\lambda_iu^\mu)\rangle \nn
&& \mbox{}
-\frac{ieF}{2} \langle\Xi(u^\mu\Q_R\lambda_i - \lambda_i\Q_Ru^\mu)\rangle 
+\frac{ieF}{4} \langle\Xi(\lambda_iu^\mu\Q_R - \Q_Ru^\mu\lambda_i)\rangle \nn
&& \mbox{}
+\frac{ieF}{4} \langle\Xi(\lambda_iu^\mu\Q_L - \Q_Lu^\mu\lambda_i)\rangle 
-\frac{eF}{2} \langle (\Q_R-\Q_L)\{\nabla^\mu\Xi,\lambda_i\}\rangle~, \\
b_i|_{eG_8^0} &=& \frac{eF}{2}\langle(\Q_R-\Q_L)\lambda_i\rangle~, \\
b_i|_{eG_8} &=& \frac{eF}{2}\langle\Xi\{\Q_R-\Q_L,\lambda_i\}\rangle~, \\
\kappa|_{e^2G_8^0} &=& \frac{e^2F^2}{2} \langle(\Q_R-\Q_L)^2\rangle~, 
\label{kappa1} \\
\kappa|_{e^2G_8} &=& 2e^2F^2 \langle\Xi(\Q_R-\Q_L)^2\rangle 
\label{kappa2}~.
\eeqa

The expressions (\ref{kappa1}) and (\ref{kappa2}) are included for
completeness only; $\kappa$ does not contribute to the order we are
concerned with.

\vspace*{2cm}


\end{document}